*Pre-print*

# TrajPy: empowering feature engineering for trajectory analysis across domains

Maurício Moreira-Soares[1,2,*], Eduardo Mossmann[3,4], Rui D. M. Travasso[5] and José Rafael Bordin[4]

[1]Oslo Centre for Biostatistics and Epidemiology, University of Oslo, Norway, [2]Centre for Bioinformatics, University of Oslo, Norway; [3]School of Engineering and Computer Science, Victoria University of Wellington, Wellington, New Zealand; [4]Department of Physics, Institute of Physics and Mathematics, Universidade Federal de Pelotas, Pelotas, Brazil. [5]CFisUC, Department of Physics, University of Coimbra, Coimbra, Portugal.

*To whom correspondence should be addressed.

**Abstract**
**Motivation:** Trajectories, sequentially measured quantities that form a path, are an important presence in many different fields, from hadronic beams in physics to electrocardiograms in medicine. Trajectory analysis requires the quantification and classification of curves either using statistical descriptors or physics-based features. To date, there is no extensive and user-friendly package for trajectory analysis available, despite its importance and potential application across domains.
**Results:** We developed a free open-source python package named TrajPy as a complementary tool to empower trajectory analysis. The package showcases a friendly graphic user interface and provides a set of physical descriptors that help characterizing these intricate structures. In combination with image analysis, it was already successfully applied to the study of mitochondrial motility in neuroblastoma cell lines and to the analysis of *in silico* models for cell migration.
**Availability:** The TrajPy package was developed in Python 3 and released under the GNU GPL-3 license. Easy installation is available through PyPi and the development source code can be found in the repository https://github.com/ocbe-uio/TrajPy/. The package release is automatically archived under the DOI 10.5281/zenodo.3656044.
**Contact:** m.m.soares@medisin.uio.no

## 1 Introduction

Trajectories are present in several fields of science with varying definitions but are intuitively understood as a set of points sequentially ordered and interconnected forming a path. More rigorously, a trajectory is defined by a sequence $(x_n(t_n))_{n\geq 0}$ of values $x$, measured at time $t_n$ with ordering index $n$

$$(x_n(t_n))_{n\geq 0} = x_0(t_0), x_1(t_1), x_2(t_2), \dots .$$

This sequence may be obtained experimentally or be the result a numerical calculation, may follow a closed mathematical form, a recursive definition, or obey a physical law or a biological mechanism (Levin, 2021). The values of $x$ are often spatial coordinates in the Euclidean space, but any quantity measured repeatedly can delineate a trajectory in an abstract space, such as blood pressure (Ji *et al.*, 2020), sunburns (Lergenmuller *et al.*, 2022) or physical activity recorded over time (Perrier *et al.*, 2022).

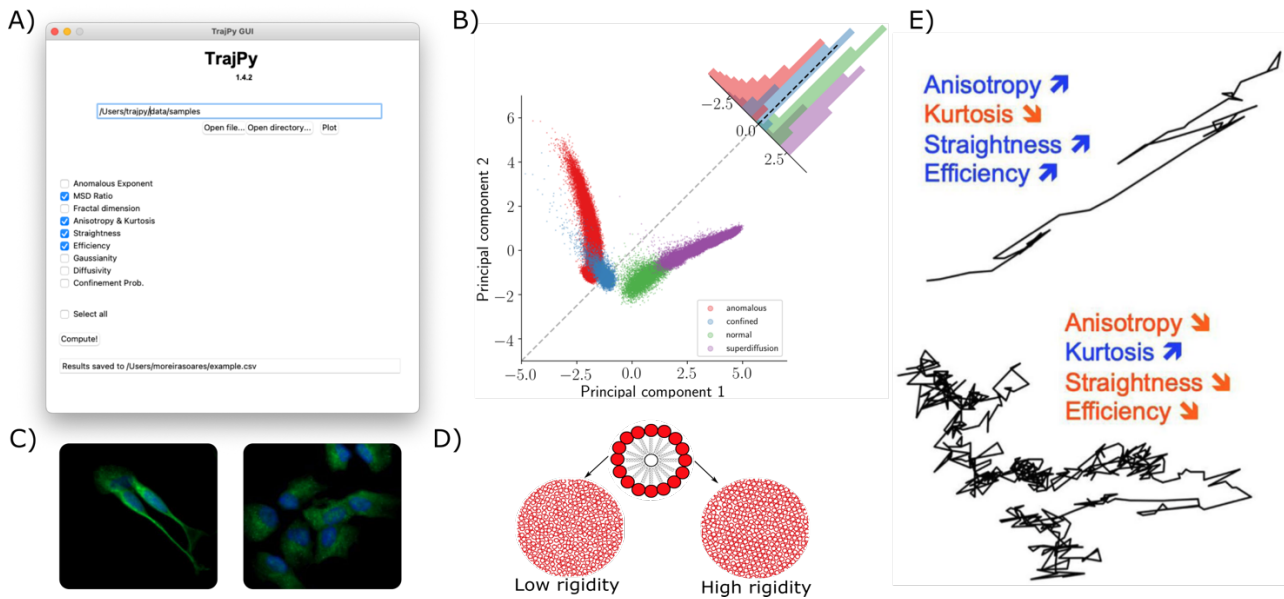

**Figure 1** – Applications and functionalities. In A) we present TrajPy's graphic user interface (GUI). In B) we show TrajPy's capabilities for trajectory classification using principal component analysis. In C) Neuroblastoma derived-cell line used for the study of mitochondrial motility and in D) a system of bead-spring polymeric entities as an *in silico* model for biological cells, both dynamics were quantified with TrajPy's features. In E) the change in a set of four features from TrajPy between two distinguished trajectories is depicted. B and E, C, and D were adapted from (Simões et al. 2021), (Soares, 2020), and (Mossman 2022), respectively.

Considerable effort has been employed to characterize complex trajectories in biology at different scales, from the diffusion of proteins and nano particles at the subcellular level (Huet *et al.*, 2006; Arcizet *et al.*, 2008), through wandering ants and migratory birds (Wolf and Wehner, 2000; Croxall *et al.*, 2005) to the study of hand tremor trajectories in Parkinson's Disease (San-Segundo *et al.*, 2020). Due to recent advances in microscopy, we can visualize the inner life of the cell, including the dynamics of mRNA, mitochondrias, microtubules, actin filaments, etc. These rich trajectory data are challenging to summarise and to model in their raw format, demanding feature extraction and quantification. In biostatistics, trajectories arise naturally in clinical trials and observational longitudinal studies, when more than two measurements are recorded for the same patient at different timepoints, such as in the studies on sunburns and physical activity previously mentioned. This type of data requires methods developed for the analysis of repeated measurements and can also benefit from feature engineering.

From the perspective of trajectory analysis in molecular dynamics (MD) simulations, among many available softwares we highlight three packages: MDAnalysis (Michaud-Agrawal *et al.*, 2011), PTraj/CPPTraj (Roe and Cheatham, 2013) and freud (Ramasubramani *et al.*, 2020). These packages are the state-of-the-art for MD analysis, but they require mastering either low-level programming or command line interfaces (CLI) which may pose a challenge for the end-user. Moreover, they are not suitable for general applications purpose.

The main goal of well-established methods such as trackpy (http://soft-matter.github.io/trackpy/) lays on image processing and these codes are heavily oriented towards specific field dependent needs. They provide a limited number of quantitative descriptors, often with focus only on the net displacement or the mean squared displacement (MSD) for estimating the diffusion exponent (Allan *et al.*, 2023). However, these measures lack sensitivity for the characterization of different kinds of trajectories in biophysics (Burnecki *et al.*, 2015). Therefore, there is a demand for specialized packages that aim to improve and democratize feature engineering for general trajectory analysis.

We propose TrajPy as a framework for tackling these challenges, aiming at broad applications across domains. The package can be integrated at the tip of image analysis pipelines, used for postprocessing of *in silico* simulations or longitudinal clinical data.

Successful modern methods perform the computation of physical properties related to the kinematics and/or morphology of the curves, to build a multidimensional space of attributes. This step permits to unveil hidden information about the trajectories by applying multivariate machine learning (ML) methods. For instance, in (Wagner *et al.*, 2017) the authors propose a set of features to quantify single cell dynamics and draw conclusions regarding the classification of cell movement. They provide a random forest classifier TraJClassifier as plugin for the image analysis software Fiji (Schindelin *et al.*, 2012). The attributes selected in this work are known in physics as good predictors for classifying movements in different types (sub-diffusion, normal diffusion, super-diffusion and anomalous diffusion), but beyond these classifications they are also helpful for identifying key differences between trajectories, even under the same diffusion regime.

We expanded the set of features proposed by Wagner et al. with fourier analysis and improved the estimation of the diffusion coefficient by implementing the Green-Kubo method. TrajPy currently offers 17 features that can be computed for any generic trajectory. It is important to note that TrajPy is not intended to replace specialized software, but it was developed as a building block that can work in synergy with other field-specific methods. In addition, it comes with a user-friendly graphical user interface (GUI) that requires no programming skills, making it accessible for experts in different fields to empower data analyses.

## 2 Methods

TrajPy is an open-source python package in continuous development on GitHub which welcome external contributions. We employ continuous integration/continuous deployment (CI/CD) with automated unit tests to assure code quality and reliability. All releases are published automatically to the PyPi repository, offering a simple method for installation and dependency management. The package development is driven by the aim of long-term maintainability and, as such, the number of external dependencies is kept at bare minimum. The core engine of the package requires only the standard packages for scientific computing *scipy* and *numpy*. In addition, to run the graphical user interface (GUI) the packages *ttkthemes* and *Pillow* are needed. Furthermore, *PyYAML* provides support for parsing molecular dynamics simulation data from LAMMPS (Plimpton *et al.*, 2021). We provide online documentation with *readthedocs*.

TrajPy consists of 3 main units of code, as described below. The heart of the package lays in *trajpy.py* and contains the class *Trajectory*, which can be initialized either as a dummy object for calling its functions, or by loading a trajectory array, or a csv trajectory file. This primary code allows the user to compute the various physical and statistical attributes such as the Mean Squared Displacement, Diffusion Coefficient and Velocity of any given trajectory (see Supplementary Table 1 for the extensive list of features). The second unit is *traj_generator.py* which consists in a collection of methods implemented to simulate different diffusion modes: confined, normal, anomalous, and direct motion. Lastly, but not least important, the *gui.py* contains the code for running the GUI, which provides a friendly interface that requires no knowledge of programming from the user (see **Figure 1**A).

We propose two independent and complementary workflows for data analysis with TrajPy. The first approach encompasses the development of a classification model for the diffusion modes aforementioned. We generate synthetic trajectories by employing 4 independent simulation engines that generate trajectories on each one of the 4 labels (sub-diffusion, normal diffusion, super-diffusion and anomalous diffusion). The space of parameters for these simulations can be explored to obtain different trajectories that obey the same diffusion regime. Then we apply feature engineering to quantify these trajectories with the proposed features in TrajPy. The data generated with the features and the labels are used to train a classifier that can be used later for classifying unseen data generated from simulations or experiments. We provide a dataset of synthetic data that can be used to train new models (Moreira-Soares, 2020). **Figure 1**B depicts the principal components for the synthetic data and the diffusion modes clusters (Soares, 2020).

The second workflow regards to statistical analysis of experimental (unlabeled) raw trajectory data. We perform the same feature engineering process on the experimental data, obtaining the same attributes used to train our classifier. Therefore, if deemed relevant, the analyst can apply the classifier to the experimental data and obtain the diffusion modes. The features can be useful to quantify different systems of interest across many areas and statistical inference can be performed to draw novel insights about the systems' nature. In addition, new classifiers can be trained based on other labels that may be interesting in other fields based on domain knowledge. For example, water quality condition affects fish trajectories, so classifying these trajectories between "Normal water quality" and "Polluted/abnormal water quality" is more relevant than diffusion modes classification in this context (Cheng *et al.*, 2019).

## 3 Validation and Results

The package was applied to study neuronal mitochondrial trafficking in neuroblastoma cell lines (Simões *et al.*, 2021). In this study, the researchers exposed the cells to mitochondrial toxins and recorded the mitochondrial trajectories using TIRF microscopy (see **Figure 1**C). By characterizing the dynamic trajectories, they analyzed how mitochondrial motility was affected. The application of TrajPy's feature engineering facilitated a deeper understanding of the underlying biological process. The findings revealed a novel quantitative approach describing how mitochondria behave in both healthy and diseased neuronal cells, demonstrating the valuable potential of TrajPy for applications in the study of subcellular dynamics.

In another study, TrajPy was employed to quantify and analyze the migration behavior of self-propelled droplets in dense fibrous media modelled *in silico* (Moreira-Soares *et al.*, 2020). By using TrajPy's feature engineering capabilities, the velocity of the cells and morphology of the cell's trajectory were measured, as a function of fiber density and adhesiveness between the cell and the matrix fibers. TrajPy enabled the comparison of simulation results with in vitro migration assay data of fibrosarcoma cells in fibrous matrices, demonstrating good agreement between the two methodologies. This study shed light on the critical role of adhesiveness in cell migration within crowded environments.

Moreover, TrajPy has been used to explore the behavior of a simplified drop-like model representing biological cells as it undergoes the jamming transition (Mossmann, Eduardo, 2022), the physical process by which viscosity increases with increasing particle density (Mongera *et al.*, 2018). In **Figure 1**D we can see the cell model with deformable boundaries and the system under two conditions with low and high rigidity. The jamming transitions have recently been recognized as key in various biological processes, including cell migration, embryo development, tissue homeostasis, and disease progression (Sadati *et al.*, 2014; Lenne and Trivedi, 2022; Oswald *et al.*, 2017; Gottheil *et al.*, 2023). By utilizing TrajPy, the behavior of the drop-like model was quantified and analyzed as pressure was increased, leading to the change in fluid viscosity. In addition, the cells trajectories were classified into the 4 diffusion modes using a classifier built with TrajPy's synthetic data. Through the application of TrajPy, the project gained valuable insights into how cell populations rapidly and significantly change their material properties during jamming transition, revealing the physiological relevance of these transition and permitting to explore potential regulatory mechanisms.

**Figure 1**E gives an intuition of how a set of 4 features implemented in TrajPy change between a trajectory with high persistency time (upper) and another with higher stochasticity (lower). More examples are provided in supplementary information, in the package's documentation and in the code repository.

## Acknowledgements

We thank all contributors from the open-source community who keep this project alive.

## Funding

MM-S received funding from the National Council for Scientific and Technological Development (CNPq - Brazil), through the proc. 235101/2014-1 and the European Union's Horizon 2020 Research and Innovation program under the Marie Skłodowska-Curie Actions Grant, agreement No. 80113 (Scientia fellowship). EHM received funding from Te Pūnaha Matatini - Centre of Excellence for Complex Systems and the Brazilian Coordination for the Improvement of Higher Education Personnel (CAPES, financing Code 001). RDMT thanks the support of FEDER funds through the Operational Programme Competitiveness Factors – COMPETE and Fundação para a Ciência e a Tecnologia through the strategic projects UIDB/04564/2020 and UIDP/04564/2020. JRB is grateful to the CNPq, proc. 403427/2021-5 and 304958/2022-0, and to the Research Support Foundation of the State of Rio Grande do Sul (FAPERGS), TO 21/2551-0002024-5, for the funding support.

## Conflict of Interest
None declared.